\documentclass[aps,twocolumn,superscriptaddress,showpacs]{revtex4}
\usepackage{epsfig}
\usepackage{graphicx}
\begin{document}
\newcommand{\bk}{{\bf k}}
\newcommand{\bK}{{\bf K}}
\newcommand{\tdia}{t^{\dagger}_{i\alpha}}
\newcommand{\tia}{t_{i\alpha}}
\newcommand{\tdja}{t^{\dagger}_{j\alpha}}
\newcommand{\tja}{t_{j\alpha}}
\newcommand{\tdjg}{t^{\dagger}_{j\gamma}}
\newcommand{\tjg}{t_{j\gamma}}
\newcommand{\tdig}{t^{\dagger}_{i\gamma}}
\newcommand{\tig}{t_{i\gamma}}
\newcommand{\tdib}{t^{\dagger}_{i\beta}}
\newcommand{\tib}{t_{i\beta}}
\newcommand{\tdjb}{t^{\dagger}_{j\beta}}
\newcommand{\tjb}{t_{j\beta}}
\newcommand{\tdnoa}{t^{\dagger}_{n+1,\alpha}}
\newcommand{\tnoa}{t_{n+1,\alpha}}
\newcommand{\tdng}{t^{\dagger}_{n\gamma}}
\newcommand{\tng}{t_{n\gamma}}
\newcommand{\tdnod}{t^{\dagger}_{n+1,\delta}}
\newcommand{\tnod}{t_{n+1,\delta}}
\newcommand{\tdag}{t^{\dagger}}
\newcommand{\al}{\alpha}
\newcommand{\be}{\beta}
\newcommand{\ca}{\gamma}
\newcommand{\de}{\delta}
\newcommand{\taud}{\tau^{\dagger}}
\newcommand{\bea}{\begin{eqnarray}}
\newcommand{\eea}{\end{eqnarray}}

\draft
\title{\bf Two-particle bound states and one-particle structure factor in a Heisenberg bilayer system
}
\author{ A. Collins, C. J. Hamer}
\affiliation{School of Physics, The University of New South Wales,
  Sydney, NSW 2052, Australia}
\date{\today}
\begin{abstract}
The $S=1/2$ Heisenberg bilayer spin model at zero temperature is studied in the dimerized phase 
using analytic triplet-wave expansions and dimer
series expansions. The occurrence of two-triplon bound states in the $S=0$ and $S=1$ channels, and antibound states in the $S=2$
channel, is predicted by the
triplet-wave theory, and confirmed by series expansions. All bound states are found to vanish at or before the
critical coupling separating the dimerized phase from the N{\' e}el phase. The critical behaviour of the total and
single-particle static transverse structure factors is also studied by series, and found to conform with theoretical
expectations. The single-particle state dominates the structure factor at all couplings.  
\end{abstract}
\pacs{PACS Indices: 05.30.-d,75.10.-b,75.10.Jm,,75.30.Kz \\
\\  \\
(Submitted to  Phys. Rev. B) }
\maketitle
\newpage

\section{INTRODUCTION}
\label{sec1}

Modern probes of material properties, such as the new inelastic neutron
scattering facilities, are reaching such unprecedented sensitivity that
they can measure the spectrum not only of a single quasiparticle
excitation, but even two-particle excitations \cite{tennant2003}. 
These quasiparticles
can collide, scatter, or form bound states just like elementary
particles in free space. The spectrum of the multiparticle excitations
is a crucial indicator of the underlying dynamics of the system.

One of the principal theoretical means of predicting the excitation
spectrum is the method of high-order perturbation series expansions 
\cite{oitmaa2006}.
We have previously used a `linked-cluster' approach to generate series
expansions for 2-particle states in 1-dimensional models \cite{trebst2000},
 but for
2-dimensional models the only high-order calculations carried out so far have been
those of Uhrig's group (e.g. \cite{knetter2000}),
 using the 
`continuous unitary
transformation' (CUTS) method, which is of only limited applicability.
One of our aims here is to extend the linked-cluster approach to
2-dimensional models, starting with the bilayer model as a simple
example.

\begin{figure}
 \includegraphics[width=0.7\linewidth]{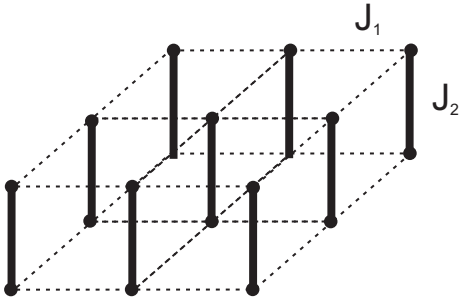}
 \caption{The bilayer Heisenberg model on a square lattice.}
 \label{fig1}
\end{figure}

The $S = 1/2$ bilayer Heisenberg antiferromagnet has attracted continuing
interest from both experimentalists and theoreticians. Experimentally, it is
of interest because many of the cuprate superconductors contain pairs of
weakly
coupled copper oxide
layers \cite{reznik1996,hayden1996,millis1996,pailhes2006}. Recently, the  
organic material 
piperazinium hexachlorodicuprate has also been found to have a bilayer
structure
\cite{stone2006}. Theoretically, it is of particular interest because it is one of the
simplest two-dimensional systems to display a dimerized, valence-bond-solid
ground state, when the interplane coupling is large. There have also
been discussions of the model in the presence of a magnetic field
\cite{sommer2001}, or doping \cite{sandvik2002,pailhes2006,zhou2007}, or
 disorder \cite{sknepnek2006}. 

The structure of the model is shown in Figure \ref{fig1}, with $S = 1/2$ spins
on the sites of the lattice, and Heisenberg antiferromagnetic couplings $J_2$
between the planes, $J_1$ within each plane:
\begin{equation}
H =  J_1 \sum_{l \ = \ 1,2} 
\sum_{<i,j>} {\bf
S_{l i} \cdot S_{l j}}
+  J_2 \sum_{ i } {\bf S_{1i} \cdot S_{2i}} 
\label{eq1}
\end{equation}
where $l = 1,2$ labels the two planes of the bilayer. The physics of the
system then depends on the coupling ratio $\lambda = J_1/J_2$. At $\lambda=0$, the
ground state consists simply of $S = 0$ dimers on each bond between the two
layers, and excitations are composed of $S = 1$ `triplon' states on one or
more bonds. At large $\lambda$, where the $J_1$ interaction is dominant, the ground
state will be a standard N{\' e}el state, with $S = 1$ `magnon' excitations.
At some intermediate critical value $\lambda_c$, a phase transition will occur
between these two phases. It is believed that this transition is of
second order, 
and is accompanied by a Bose-Einstein condensation of
triplons/magnons in the ground state. 

Theorists have discussed this model using series expansion methods 
\cite{hida1992,gelfand1996,zheng1997},
quantum Monte Carlo )QMC) simulations at small temperatures 
\cite{sandvik1994,sandvik1995,sandvik1996}, Schwinger-boson mean-field
theory \cite{millis1993,miyazaki1996}, and spin-wave theory
\cite{matsuda1990,chubukov1995,kotov1998,shevchenko1999}.
The QMC analysis of Sandvik and Scalapino \cite{sandvik1996} found the transition at $\lambda_c = 0.398(3)$, with a
critical index $\nu ]simeq 0.7$.
in agreement with the O(3)
nonlinear sigma model prediction, while the 
  exponent-biased series analysis
of Zheng \cite{zheng1997} put the critical point at $\lambda_c = 0.394(1)$.
Early spin-wave estimates \cite{chubukov1995} were well away from this position, but
the improved Brueckner approach of Sushkov {\it et al.} 
\cite{kotov1998,shevchenko1999} gave a
remarkably accurate estimate of the critical point and critical index, 
and also the 1-particle
dispersion in the model.

Our particular aim here is to study the two-triplon states within the
dimerized regime, with particular emphasis on the occurrence of bound states,
and to explore their behaviour in the vicinity of the critical point.
The two-particle bound states can give important insights into the dynamical
behaviour of the model. It is also possible that they may be detected experimentally at the
new generation of inelastic neutron scattering facilities, or by other means.

We use two methods to investigate the two-particle states. A modified
triplet-wave approach, described in Section \ref{sec2}, gives a qualitative picture of
these states, valid at small couplings $\lambda$. Series expansion calculations,
sketched in Section \ref{sec3}, are
then used to obtain more accurate results, and to explore the behaviour near
the critical point.
Series expansions are also presented for the single-particle and total transverse structure factors.
Our conclusions are summarized in Section \ref{sec5}. 

\section{Modified triplet-wave theory.}
\label{sec2}

Analogues of spin-wave theory in a dimerized phase have been discussed by
several authors. 
Sachdev and Bhatt \cite{sachdev1990}
used a `bond-operator' representation to describe the dimers and their
spin-triplet excitations, which employed both triplet and singlet operators,
with a constraint between them to ensure that no two triplets can occupy the
same site. The constraint is awkward to implement, and so Kotov {\it et al.}
\cite{kotov1998} discarded the singlet operator, and replaced it by an infinite on-site
repulsion between triplets, implemented via  a self-consistent Born approximation,
valid when the density of triplets is low. We have presented an alternative
approach \cite{collins2006}, where the exclusion constraint is implemented automatically
by means of projection operators. The absence of any constraint makes the
formalism easier and more transparent to apply, but at the price of extra
many-body interaction terms.  
This is the method used here.

The Hamiltonian for the Heisenberg bilayer systemn can be rewritten
\begin{equation}
H = \sum_{ i } {\bf S_{1i} \cdot S_{2i}} + \lambda \sum_{1  =  1,2} 
\sum_{<i,j>} {\bf
S_{\lambda i} \cdot S_{\lambda j}}
\label{eq10}
\end{equation}
For $\lambda = 0$, the system reduces to independent dimers as shown in
Figure \ref{fig1}. 
Let us consider a single dimer with two spins ${\bf S_1,S_2}$.
The four states in the Hilbert space consist of a singlet and three
triplet states with total spin $S=0,1$ respectively, and eigenvalues
\begin{eqnarray}
{\bf S_1 \cdot S_2} = \left\{ \begin{array}{cc}
-3/4 & (S=0) \\
+1/4 & (S=1)
\end{array}
\right.
\label{eq11}
\end{eqnarray}
We denote the singlet ground state as $|0 \rangle$, and introduce triplet
creation operators that create the triplet states out of the vacuum
$|0 \rangle$, as follows
\begin{eqnarray}
|0 \rangle & = & \frac{1}{\sqrt{2}}[|\uparrow \downarrow \rangle -|\downarrow
\uparrow \rangle] \nonumber \\
|1,x \rangle & = & t^{\dagger}_x|0 \rangle = -\frac{1}{\sqrt{2}}[|\uparrow \uparrow \rangle -|\downarrow
\downarrow \rangle] \nonumber \\
|1,y \rangle & = & t^{\dagger}_y|0 \rangle = \frac{i}{\sqrt{2}}[|\uparrow \uparrow \rangle +|\downarrow
\downarrow \rangle] \nonumber \\
|1,z \rangle & = & t^{\dagger}_z|0 \rangle = \frac{1}{\sqrt{2}}[|\uparrow \downarrow \rangle +|\downarrow
\uparrow \rangle]
\label{eq12}
\end{eqnarray}
Then the spin operators ${\bf S_1}$ and ${\bf S_2}$ can be represented
in terms of triplet operators by
\begin{eqnarray}
S_{1\alpha} & = &
\frac{1}{2}[t^{\dagger}_{\alpha}(1-t^{\dagger}_{\gamma}t_{\gamma}) +
(1-t^{\dagger}_{\gamma}t_{\gamma}) t_{\alpha}
-i\epsilon_{\alpha\beta\gamma}t^{\dagger}_{\beta}t_{\gamma}]
 \nonumber \\
S_{2\alpha} & = &
\frac{1}{2}[-t^{\dagger}_{\alpha}(1-t^{\dagger}_{\gamma}t_{\gamma}) -
(1-t^{\dagger}_{\gamma}t_{\gamma}) t_{\alpha}
\nonumber \\
& &   -i\epsilon_{\alpha\beta\gamma}t^{\dagger}_{\beta}t_{\gamma}]
\label{eq13}
\end{eqnarray}
where $\alpha,\beta,\gamma$ take the values $x,y,z$ and repeated indices
are summed over. This is similar to the representation of Sachdev and
Bhatt \cite{sachdev1990}, except that we have omitted singlet operators
$s^{\dagger},s$, but used projection operators
$(1-t^{\dagger}_{\gamma}t_{\gamma})$ instead. Assume the triplet
operators obey bosonic commutation relations
\begin{equation}
[t_{\alpha},t^{\dagger}_{\beta}] = \delta_{\alpha\beta},
\label{eq14}
\end{equation}
then one can show that within the physical subspace (i.e. total number
of triplet states is 0 or 1), the representation (\ref{eq13}) obeys the
correct spin operator algebra
\begin{equation}
[S_{1\alpha},S_{1\beta}]  =  i\epsilon_{\alpha\beta\gamma}S_{1\gamma},
\hspace{5mm} [S_{2\alpha},S_{2\beta}] =
i\epsilon_{\alpha\beta\gamma}S_{2\gamma},
\nonumber
\end{equation}

\begin{equation}
[S_{1\alpha},S_{2\beta}]  =  0
\label{eq15b}
\end{equation}
\begin{equation}
{\bf S}_1^2  =  {\bf S}_2^2 = 3/4, \hspace{5mm} {\bf S_1 \cdot S_2} =
t^{\dagger}_{\alpha}t_{\alpha} - 3/4
\label{eq15c}
\end{equation}
The projection operators ensure that we remain within the subspace.

Returning to the bilayer system, we can now define triplet operators
$t^{\dagger}_{n\alpha},t_{n\alpha}$ for each dimer $n$ in the system.
For a system of $N$ dimers, the Hamiltonian now can be expressed in terms
of triplet operators as
\begin{widetext}
\begin{eqnarray}
H & = &  -\frac{3N}{4} + \sum_n t^{\dagger}_{n\alpha}t_{n\alpha}
+\frac{\lambda}{2} \sum_{<ij>} \{ \tdia \tdja + \tia \tja +\tia \tdja + \tdia
\tja \}
-\frac{\lambda}{2} \sum_{<ij>} \{
(\tdia \tdib \tib + \tdib \tib \tia) (\tdja +\tja) \nonumber \\
 & &  +(\tdia + \tia) 
(\tdja \tdjb \tjb + \tdjb \tjb \tja) + \tdia \tib \tdja \tjb - \tdia \tib
\tdjb \tja \} \nonumber \\
 & & 
+\frac{\lambda}{2} \sum_{<ij>} \{
(\tdia \tdib \tib + \tdib \tib \tia) (\tdja \tdjg \tjg + \tdjg \tjg
\tja) \}
\label{eq16}
\end{eqnarray}
\end{widetext}
This expression includes terms containing up to 6 triplet operators.

Next, perform a Fourier transform
\begin{eqnarray}
t_{{\bf k}\al} & = & (\frac{1}{N})^{1/2} \sum_{\bf n} e^{i{\bf k.n}} t_{{\bf
n}\al} \nonumber \\
t^{\dagger}_{{\bf k}\al} & = & (\frac{1}{N})^{1/2} \sum_{\bf n} e^{-i{\bf
k.n}} \tdag_{{\bf n}\al}
\label{eq17}
\end{eqnarray}
(we set the spacing between dimers $d = 1$),
then the Hamiltonian becomes
\begin{widetext}
\begin{eqnarray}
H & = & -\frac{3N}{4} + \sum_{\bk} \tdag_{{\bk}\al} t_{\bk \al}
+\lambda \sum_{\bk} \gamma_{\bk} [ \tdag_{\bk \al} \tdag_{-\bk \al} + t_{\bk
\al} t_{-\bk \al} +2 \tdag_{\bk \al} t_{\bk \al} ]
 \nonumber \\
 & &
-\frac{\lambda}{N} \sum_{1234} \{  \delta_{1+2+3-4} [ (\tdag_{1 \al} \tdag_{2
\al} \tdag_{3\ca} t_{4 \ca} + \tdag_{4 \ca} t_{3 \ca} t_{2 \al} t_{1
\al}) 
 (\gamma_1 + \gamma_2)] +
\delta_{1+2-3-4}[\tdag_{1\al}\tdag_{2\ca}t_{3\al}t_{4\ca}(\gamma_1+\gamma_2+\gamma_3+\gamma_4)
 \nonumber \\
 & &   
 + \gamma_{{\bf 1-3}} \tdag_{1 \be} \tdag_{2 \be} t_{3 \ca}
t_{4 \ca} - \gamma_{{\bf 1-4}}\tdag_{1 \be} \tdag_{2 \ca} t_{3 \be} t_{4 \ca}) ]
\}
 \nonumber \\
 & &
+ \frac{\lambda}{N^2} \sum_{1-6} \{ \delta_{1+2+3+4-5-6} [ \gamma_{{\bf 3+4-6}}
(\tdag_{1 \al} \tdag_{2 \ca} \tdag_{3 \al} \tdag_{4 \be} t_{5 \ca} t_{6
 \be} + \tdag_{6 \be} \tdag_{5 \ca} t_{4 \be} t_{3 \al} t_{2 \ca} t_{1
\al}) ]  \nonumber \\
 & &  + \delta_{1+2+3-4-5-6}
   \tdag_{1 \al} \tdag_{2 \be} \tdag_{3 \ca}
t_{4 \ca} t_{5 \be} t_{6 \al}(\gamma_{{\bf 3-4-6}} + \gamma_{{\bf 2+3-4}}) 
\}
\label{eq18}
\end{eqnarray}
\end{widetext}
where the indices $1 - 6$ are shorthand for momenta ${\bf k_1} - {\bf
k_6}$, and
\begin{equation}
\gamma_{\bf k} = \frac{1}{2}(\cos k_x + \cos k_y)
\end{equation}
for the square lattice.
Henceforward, we drop the 6-particle terms.

Finally, as in a standard spin-wave analysis, we perform a Bogoliubov
transform
\begin{equation}
t_{\bk\al} = c_{\bk}\tau_{\bk\al} + s_\bk \tau^{\dagger}_{-\bk\al}
\label{eq19}
\end{equation}
where $c_\bk = \cosh \theta_\bk$, $s_\bk = \sinh \theta_\bk$, $\theta_{-\bk} =
\theta_\bk$, which preserves the boson commutation relations
\begin{equation}
[\tau_{\bk\al},\taud_{\bk'\beta}] = \delta_{\bk\bk'}\delta_{\al\be}
\label{eq20}
\end{equation}
and is intended to diagonalize the Hamiltonian up to quadratic terms.
After normal ordering, the transformed Hamiltonian up to fourth order
terms reads
\begin{equation}
H = W_0 + H_2 + H_3 + H_4,
\label{eq21}
\end{equation}
where the constant term is
\begin{widetext}
\begin{eqnarray}
W_0 & = & 3N\left\{  -\frac{1}{4}+R_2 +
2\lambda(R_3+R_4)-2\lambda[2(R_3+R_4)(R_1+4R_2) 
  +\frac{1}{N^2}\sum_{12}
\gamma_{{\bf 1-2}}(c_1s_1c_2s_2-s_1^2s_2^2)] \right. \nonumber \\
 & & \left. +2\lambda[(R_3+R_4)(R_1+4R_2)^2
  +\frac{1}{N^3}\sum_{123}\gamma_{{\bf 1+2-3}}(c_1s_1(4c_2s_2c_3s_3+
6c_2s_2s_3^2+6s_2^2s_3^2)+4s_1^2s_2^2s_3^2)] \right\}
\label{eq22}
\end{eqnarray}
\end{widetext}
expressed in terms of the momentum sums
\begin{eqnarray}
R_1 & = & \frac{1}{N} \sum_{\bk} c_{\bk} s_{\bk} \nonumber \\
R_2 & = & \frac{1}{N} \sum_{\bk}  s_{\bk}^2
\nonumber \\
R_3 & = & \frac{1}{N} \sum_{\bk} c_{\bk} s_{\bk} \gamma_{\bk}
\nonumber \\
R_4 & = & \frac{1}{N} \sum_{\bk} s_{\bk}^2 \gamma_{\bk}.
\label{eq23}
\end{eqnarray}

The quadratic terms are
\begin{equation}
H_2 = \sum_{\bk,\al} [E_{\bk} \taud_{\bk\al} \tau_{\bk\al} +Q_\bk(\tau_{\bk\al}\tau_{-\bk\al} + \taud_{\bk\al} \taud_{-\bk\al})]
\label{eq24}
\end{equation}
where
\begin{widetext}
\begin{eqnarray}
E_{\bk}  & =  & (c_{\bk}^2 + s_{\bk}^2)(1+2\lambda\gamma_{\bk})+4\lambda\gamma_{\bk}c_{\bk}
s_{\bk} \nonumber \\
 & & -\lambda[4(c_{\bk}^2+s_{\bk}^2)(\gamma_{\bk}
(R_1+4R_2) + 4(R_3+R_4)
 -\frac{1}{N}\sum_1s_1^2\gamma_{{\bf k-1}})
\nonumber \\ & &
  +8c_{\bk}s_{\bk}
(\gamma_{\bk}
(R_1+4R_2) + (R_3+R_4)
  +\frac{1}{N}\sum_1c_1s_1\gamma_{{\bf
k-1}})] 
\label{eq25}
\end{eqnarray}
\begin{eqnarray}
Q_{\bk} & = & c_{\bk}s_{\bk}(1
+2\lambda\gamma_{\bk})+\lambda\gamma_{\bk}(c_{\bk}^2+s_{\bk}^2) \nonumber \\
 & & -\lambda[2(c_{\bk}^2+s_{\bk}^2)(\gamma_{\bk}
(R_1+4R_2)+(R_3+R_4)
+\frac{1}{N}\sum_1
c_1s_1\gamma_{\bk-{\bf
1}}) \nonumber \\
 & &+4c_{\bk}s_{\bk}(\gamma_{\bk} 
(R_1+4R_2)+4(R_3+R_4)
 -\frac{1}{N}\sum_1
s_1^2\gamma_{\bf k-1})]
\label{eq26}
\end{eqnarray}
The fourth-order terms are
\begin{eqnarray}
H_4 & = & -\frac{\lambda}{N}\sum_{1234}[\delta_{1+2+3+4}\Phi^{(1)}_4
(\taud_{1\al}\taud_{2\al}\taud_{3\ca}\taud_{4\ca} +
\tau_{1\al}\tau_{2\al}\tau_{3\ca}\tau_{4\ca}) +
\delta_{1+2-3-4}(\Phi^{(2)}_4
\taud_{1\al}\taud_{2\al}\tau_{3\ca}\tau_{4\ca}
+\Phi_4^{(3)}\taud_{1\al}\taud_{2\ca}\tau_{3\al}\tau_{4\ca})
\nonumber \\
& & +\delta_{1+2+3-4}\Phi^{(4)}_4
(\taud_{1\al}\taud_{2\al}\taud_{3\ca}\tau_{4\ca} +
\taud_{4\ca}\tau_{3\ca}\tau_{2\al}\tau_{1\al})]
\label{eq28}
\end{eqnarray}
\end{widetext}
where we have used the shorthand notation $1 \cdots 4$ for momenta $k_1 \cdots k_4$, and the vertex functions
$\Phi^{(i)}_4$ are listed in Appendix A.
These results were obtained or confirmed using a symbolic manipulation program
written in PERL.

The condition that the off-diagonal quadratic terms vanish is
\begin{equation}
Q_{\bk} =0.
\label{eq29}
\end{equation}
In a conventional spin-wave approach, this would be implemented in leading order only, giving the condition
\begin{equation}
\tanh 2\theta_{\bk} = \frac{2s_{\bk}c_{\bk}}{c_{\bk}^2+s_{\bk}^2} =
 -\frac{2\lambda\gamma_{\bk}}{[1+2\lambda\gamma_{\bk})}
\label{eq30}
\end{equation}
This would leave some residual off-diagonal quadratic terms, arising from the normal-ordering of quartic operators. In a
`modified' approach \cite{gochev1994}, we demand that these terms vanish entirely up to the order calculated, giving the
modified condition

\begin{widetext}
\begin{equation}
\tanh 2\theta_{\bk}  = 
 -\frac{2\lambda[\gamma_{\bk}
-2(\gamma_{\bk}(R_1+4R_2)+(R_3+R_4)
+\frac{1}{N}\sum_1 c_1s_1 \gamma_{{\bf k-1}})]
}{[1+2\lambda(\gamma_{\bk}-2(\gamma_{\bk}(R_1+4R_2)+4(R_3+R_4)
-\frac{1}{N}\sum_1s_1^2\gamma_{{\bf
k-1}}))]}
\label{eq31}
\end{equation}
\end{widetext}
Self-consistent solutions for the N equations (\ref{eq31}), with the four parameters $R_1 \cdots R_4$ given by equation
(\ref{eq23}), can easily be found by numerical means, starting from the conventional result (\ref{eq30}).

\subsection{Expansion in powers of $\lambda$}
\label{sec2a}

As a first check on the formalism, one may calculate the leading terms
in an expansion of the energy eigenvalues in powers of $\lambda$.
Solving the modified equation (\ref{eq31}) self-consistently
to order $\lambda^2$, we find
\begin{eqnarray}
s_{\bk} & = & 
-\lambda\gamma_{\bk} + \frac{\lambda^2}{2}(4\gamma_{\bk}^2
-\gamma_{\bk}-1)
\nonumber \\
c_{\bk} & = & 1 + \frac{1}{2}\lambda^2\gamma_{\bk}^2
\label{eq32}
\end{eqnarray}
with the lattice sums (\ref{eq23}) 
\begin{eqnarray}
R_1 & = & O(\lambda^4), \hspace{5mm} R_2 = \frac{\lambda^2}{4}+\frac{\lambda^3}{4}
+O(\lambda^4), \nonumber \\
R_3 & = & -\frac{\lambda}{4} - \frac{\lambda^2}{8}+O(\lambda^3), \hspace{5mm} R_4 =
O(\lambda^3)
\label{eq33}
\end{eqnarray}
The leading-order behaviour of the vertex functions may easily be
deduced from Appendix A.

\begin{figure}
 \includegraphics[width=0.7\linewidth]{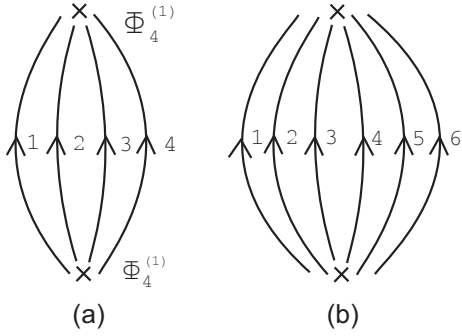}
 \caption{Perturbation diagrams contributing to the ground-state energy.}
 \label{fig2}
\end{figure}

Substituting in equation (\ref{eq22}), the ground state energy per dimer 
is
\begin{eqnarray}
\epsilon_0 & = &  \frac{W_0}{N} 
  \sim 
-3[\frac{1}{4}+\frac{\lambda^2}{4}+\frac{\lambda^3}{8}]
\hspace{5mm} \lambda \rightarrow 0
\label{eq34}
\end{eqnarray}
in agreement with dimer series expansion results previously obtained for this
model \cite{zheng1997}. One can easily show that perturbation diagrams
such as those in Figure \ref{fig2} do not contribute until
$O(\lambda^4)$ or higher.

The energy gap at leading order can be found from equation (\ref{eq25}):
\begin{equation}
E_k \sim 1
+2\lambda\gamma_{\bk} +4\lambda^2-2\lambda^2\gamma_{\bk}^2
 \hspace{5mm} \lambda \rightarrow 0
\label{eq35}
\end{equation}
Note that in linear spin-wave theory, when $\tanh 2\theta_{\bk}$ is given
by (\ref{eq30}) and the energy gap is given by the first line of equation
(\ref{eq25}), the energy gap is
\begin{equation}
E_k = \sqrt{1+4\lambda\gamma_{\bk}}
\label{eq35a}
\end{equation}
which vanishes at $\lambda = 1/4, \gamma_{\bk} = -1$, i.e. ${\bf k} =
(\pi,\pi)$. This marks a phase transition with critical index $\nu =
1/2$, and the end of the dimerized
phase, in this approximation.

\begin{figure}
\includegraphics[width=0.7\linewidth]{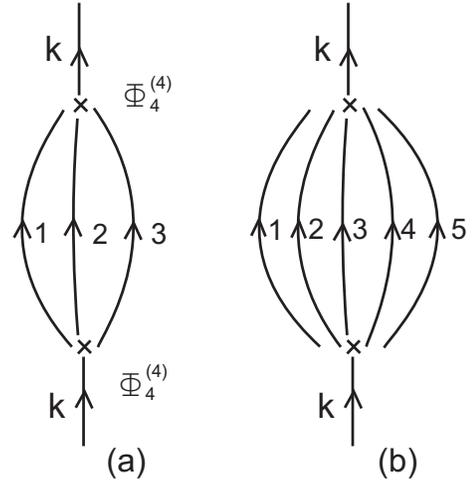}
 \caption{Perturbation diagrams contributing to the one-particle energy.}
 \label{fig3}
\end{figure}

The perturbation diagram Figure \ref{fig3}a) also
contributes to the energy gap at order $\lambda^2$.
Note that diagram \ref{fig3}a) does not appear in the formalism of Shevchenko
{\it et al.} \cite{shevchenko1999,kotov1998}; the extra terms in our formalism are needed to implement the
hardcore constraint that two triplons cannot occupy the same site.
At leading order, the contribution of this diagram
is
\begin{equation}
\Delta E_k^{3a)} \sim -2\lambda^2
 \hspace {5mm} \lambda \rightarrow 0
\label{eq36}
\end{equation}
(see the next section for further details). This gives a total
single-particle energy
\begin{equation}
\epsilon_k \sim 
1+2\lambda\gamma_{\bk}+2\lambda^2(1-\gamma_{\bk}^2),
\hspace{5mm} \lambda \rightarrow
0
\label{eq38}
\end{equation}
which again agrees with series expansion results \cite{zheng1997}.

The minimum energy gap lies at ${\bf k} = (\pi,\pi)$. If we compare equation 
(\ref{eq38}) at small momentum ${\bf p} = (\pi,\pi)-{\bf k}$ with the
continuum dispersion relation for a free boson,

\begin{equation}
\epsilon_k \sim \sqrt{m^2c^4 + p^2c^2}
\label{eq39}
\end{equation}
we readily discover the leading behaviour of the effective triplon
parameters, i.e. the triplon mass

\begin{equation}
m \sim \frac{1}{\lambda}[1-2\lambda+O(\lambda^2)]
\label{eq40}
\end{equation}
and the `speed of light' or triplon velocity

\begin{equation}
c^2 \sim 
\lambda+O(\lambda^3)
\label{eq41}
\end{equation}
in lattice units. Note that the mass diverges and the speed of light
vanishes as $\lambda \rightarrow 0$.

\subsection{Numerical Results}
\label{sec2b}

Writing the Hamiltonian as

\begin{equation}
H = H_0 + V
\label{eq42}
\end{equation}
where

\begin{equation}
H_0 = W_0 + H_2
\label{eq43}
\end{equation}
and
\begin{equation}
V = H_4 
\label{eq44}
\end{equation}
(6-particle terms being neglected)
we can treat $H_0$ as the unperturbed Hamiltonian and $V$ as a
perturbation to obtain the leading-order corrections to the predictions
for physical quantities outlined in the previous section.
Numerical results for the model have been obtained using the
finite-lattice method. The momentum sums are carried out for a fixed
lattice size $L \times L = N$,
 using corresponding discrete values for the momentum
$\bk$, e.g.

\begin{eqnarray}
k_x & = & \frac{2\pi n}{L}, \hspace{5mm}n=1, \cdots L
\nonumber \\
k_y & = & \frac{2\pi m}{L}, \hspace{5mm}m=1, \cdots L
\label{eq45}
\end{eqnarray}
Results were obtained for $L$ up to 100.

\subsubsection{Ground-state energy}
\label{seca1}

\begin{figure}
 \includegraphics[width=1.0\linewidth]{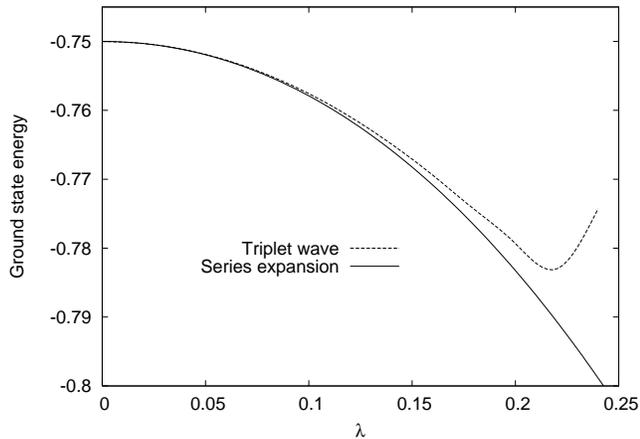}
 \caption{Ground-state energy per dimer as a function of $\lambda$. The solid
line
is the estimate from series expansions, and the dashed line is the
triplet-wave estimate.
}
 \label{fig4}
\end{figure}

The leading correction to the ground-state energy corresponds to the
diagram in Figure \ref{fig2}a). Its
contribution is

\begin{widetext}
\begin{eqnarray}
\Delta \epsilon_0^{2a)}  & = &  \frac{-3\lambda^2}{N^3}\sum_{1234}
\delta_{1+2+3+4}\frac{\Phi_4^{(1)}(1234)}{(E_1+E_2+E_3+E_4)}
  \left[3\Phi_4^{(1)}(1234)
  +\Phi_4^{(1)}(1324)
  +\Phi_4^{(1)}(1423)\right]
\label{eq45b}
\end{eqnarray}
\end{widetext}
In leading order one can show that this term is $O(\lambda^4)$,
whereas diagrams such as Figure \ref{fig2}b) are $O(\lambda^6)$ or
higher.
Figure
\ref{fig4} shows the behaviour of the ground-state energy as a function
of $\lambda$ resulting from this modified triplon theory, as compared
with the high-order dimer series calculations of Zheng 
 \cite{zheng1997}.
It can be seen that out to $\lambda \simeq 0.1$ there is quantitative agreement between our calculation
and the series estimates, but some discrepancy emerges at larger $\lambda$.

\subsubsection{One-particle spectrum}
\label{sec2a2}

The leading correction to the one-particle spectrum corresponds to the
diagram in Figure \ref{fig3}a). Its contribution
is

\begin{widetext}
\begin{eqnarray}
\Delta E_k^{3a)} & = & \frac{2\lambda^2}{N^2}\sum_{ 123}\delta_{
{\bf 1+2+3-k}}\frac{\Phi_4^{(4)}(
123k)}{(E_k-E_1-E_2-E_3)}
\left[3\Phi_4^{(4)}(123k)
+\Phi_4^{(4)}(321k)
+\Phi_4^{(4)}(312k)
\right]
\label{eq47}
\end{eqnarray}
\end{widetext}
In leading order, this term is $O(\lambda^2)$, as stated in the
previous section, while diagrams like \ref{fig3}b) are
$O(\lambda^4)$ or higher.

\begin{figure}
 \includegraphics[width=1.0\linewidth]{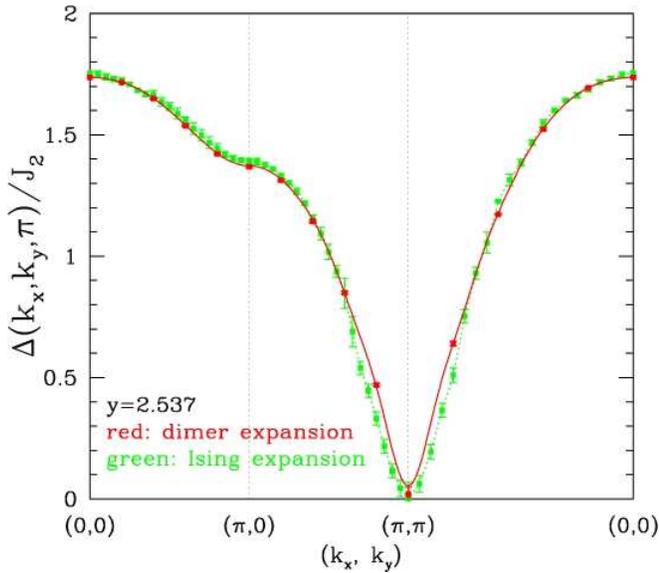}
 \caption{[Colour online] One-particle dispersion relation at the critical point ($y =
1/\lambda$), as estimated from both dimer (solid line) and Ising (dashed line) expansions
\cite{zheng1997}.}
 \label{fig5}
\end{figure}

The dispersion of the one-particle energy as a function of momentum ${\bf
k}$ at the critical point is
illustrated in Figure \ref{fig5}, as estimated from two different  series 
expansions
by Zheng \cite{zheng1997}. It can be seen that the two expansions agree
well at the critical point, and that the energy gap vanishes there at
the N{\' e}el point ${\bf k} = (\pi,\pi)$.

\begin{figure}
 \includegraphics[width=1.0\linewidth]{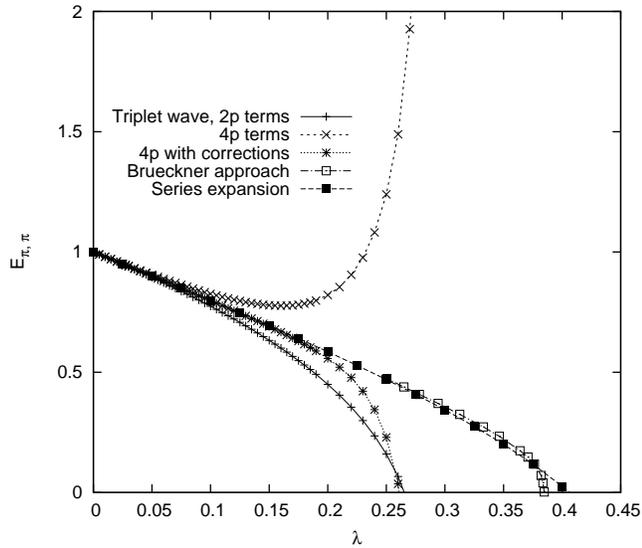}
 \caption{ Energy gap at ${\bf k} = (\pi,\pi)$ as a function of 
$\lambda$. The solid squares show the series estimates
\cite{zheng1997},
and the open squares are 
results from Shevchenko {\it et al.} 
\cite{shevchenko1999}, 
 while the stars show the improved 
triplet-wave results. The contributions from 2-triplon and 4-triplon
terms are shown separately. 
}
 \label{fig6}
\end{figure}

The triplet-wave and series estimates of the energy gap at 
${\bf k} = (\pi,\pi)$ are
compared in Figure \ref{fig6}.
 It
can be seen that the inclusion of the corrections from diagram \ref{fig3}a)
improves the agreement with series 
substantially, bringing quantitative agreement out to $\lambda \simeq
0.15$. Beyond that, the triplet-wave estimates begin to diverge, as
higher-order contributions become more important. 
The self-consistent Born approach of Kotov et al. \cite{kotov1998,shevchenko1999} is more accurate than our
approach at large $\lambda$; but neither approach can compete with
series methods for accuracy. Our object here mainly is to understand the
qualitative behaviour of the model.

\subsubsection{Two-triplon bound states}
\label{sec2a3}

It has been found in previous studies of dimerized
antiferromagnetic systems in one dimension \cite{uhrig1996,shevchenko1999} that the 
quartic terms in
the Hamiltonian lead to attraction between two elementary triplons,
giving rise to $S=0$ and $S=1$ bound states. We look for solutions of
the two-body Schr{\" o}dinger equation

\begin{equation}
H|\psi \rangle = E|\psi \rangle.
\label{eq48}
\end{equation}

The two-body wave functions $|\psi({\bf K}) \rangle$ can be written as follows:
\begin{flushleft}
{\it Singlet sector ($S=0$):}
\end{flushleft}
\begin{equation}
|\psi^S({\bf K}) \rangle =
\frac{1}{\sqrt{6}}\sum_{{\bf q},\alpha}\psi^S({\bf K},{\bf q})\tau^{\dagger}_{{\bf
K}/2+{\bf q},\alpha}
\tau^{\dagger}_{{\bf K}/2-{\bf q},\alpha}|0 \rangle
\label{eq49}
\end{equation}
where ${\bf K}$ is the centre-of-mass momentum and ${\bf q}$ the relative momentum
of the two particles,
and the scalar wave function is symmetric,
\begin{equation}
\psi^S({\bf K},-{\bf q})=\psi^S({\bf K},{\bf q})
\label{eq53}
\end{equation}

\begin{flushleft}
{\it Triplet sector ($S=1$):}
\end{flushleft}
\begin{equation}
|\psi^T_{\alpha}({\bf K}) \rangle =
\frac{1}{2}\sum_{{\bf q},\beta,\gamma}\epsilon_{\alpha\beta\gamma}
\psi^T({\bf K},{\bf q})\tau^{\dagger}_{{\bf K}/2+{\bf q},\beta}\tau^{\dagger}_{{\bf
K}/2-{\bf q},\gamma}|0 \rangle
\label{eq50}
\end{equation}
with the wave function antisymmetric
\begin{equation}
\psi^T({\bf K},-{\bf q})=-\psi^T({\bf K},{\bf q}).
\label{eq56}
\end{equation}
We will not write out the quintuplet states explicitly.

From equation (\ref{eq48}) one can readily derive the integral
Bethe-Salpeter equation satisfied by the bound-state wave functions:

\begin{eqnarray}
[E^{S,T,Q}({\bf K})-E_{{\bf K}/2+{\bf q}}-E_{{\bf K}/2-{\bf q}}]\psi^{S,T,Q}({\bf K},{\bf
q})
= \nonumber \\
\frac{1}{N}\sum_{{\bf p}}M^{S,T,Q}({\bf K},{\bf q},{\bf p})\psi^{S,T,Q}({\bf K},{\bf
p})
\label{eq51}
\end{eqnarray}
in each sector S,T or Q.

\begin{figure}
 \includegraphics[width=0.8\linewidth]{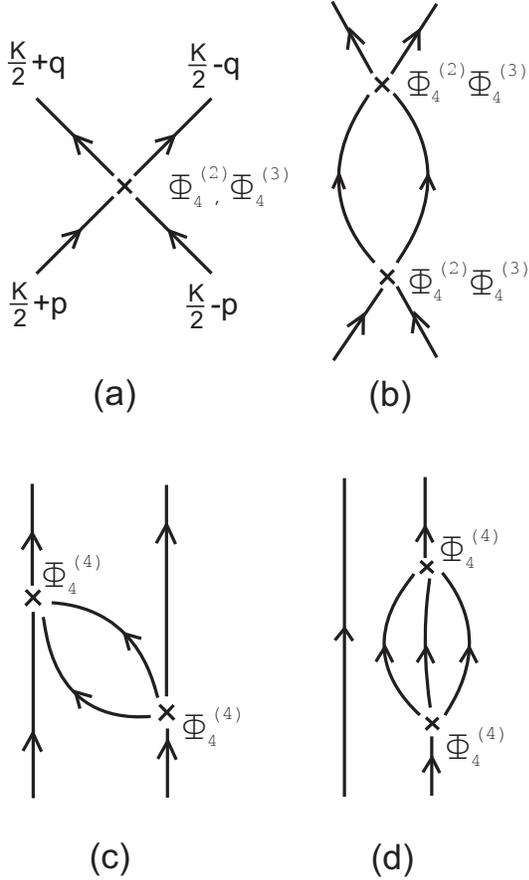}
 \caption{Perturbation diagrams contributing to the 2-particle
scattering amplitude.}
 \label{fig7}
\end{figure}

In leading order, the scattering amplitudes $M^{S,T,Q}({\bf K},{\bf q},{\bf
p})$ are simply
given by the 4-particle vertex from the perturbation operator $V$, Figure
\ref{fig7}a). Hence we find for the different sectors:

\begin{widetext}
\begin{equation}
M^S({\bf K},{\bf q},{\bf p})  =
-2\lambda[3\Phi^{(2)}_4({\bf K}/2+{\bf q},{\bf K}/2-{\bf q},{\bf K}/2+{\bf
p},{\bf
K}/2-{\bf p})
  +\Phi^{(3)+}_4({\bf {\bf K}}/2+{\bf q},{\bf {\bf K}}/2-{\bf q},{\bf K}/2+
{\bf p},{\bf K}/2-{\bf
p})]
\label{eq52}
\end{equation}
\begin{equation}
M^T({\bf K},{\bf q},{\bf p}) =
-2\lambda\Phi^{(3)-}_4({\bf K}/2+{\bf q},{\bf K}/2-{\bf q},{\bf K}/2+{\bf p},{\bf
K}/2-{\bf p})
\label{eq55}
\end{equation}
\begin{equation}
M^Q({\bf K},{\bf q},{\bf p}) =
-2\lambda\Phi^{(3)+}_4({\bf K}/2+{\bf q},{\bf K}/2-{\bf q},{\bf K}/2+{\bf
p},{\bf
K}/2-{\bf p})
\label{eq57}
\end{equation}
\end{widetext}
where the wave function is once again symmetric
in the quintuplet sector
\begin{equation}
\psi^Q({\bf K},-{\bf q})=\psi^Q({\bf K},{\bf q}).
\label{eq58}
\end{equation}
and the symmetric and antisymmetric pieces of the vertex function
$\Phi_4^{(3)}$ are defined:
\begin{equation}
\Phi^{(3)\pm}_4  \equiv  \frac{1}{2}[\Phi^{(3)}_4(1234)\pm
\Phi^{(3)}_4(1243)].
\label{eq54}
\end{equation}

\begin{figure}
 \includegraphics[width=0.9\linewidth]{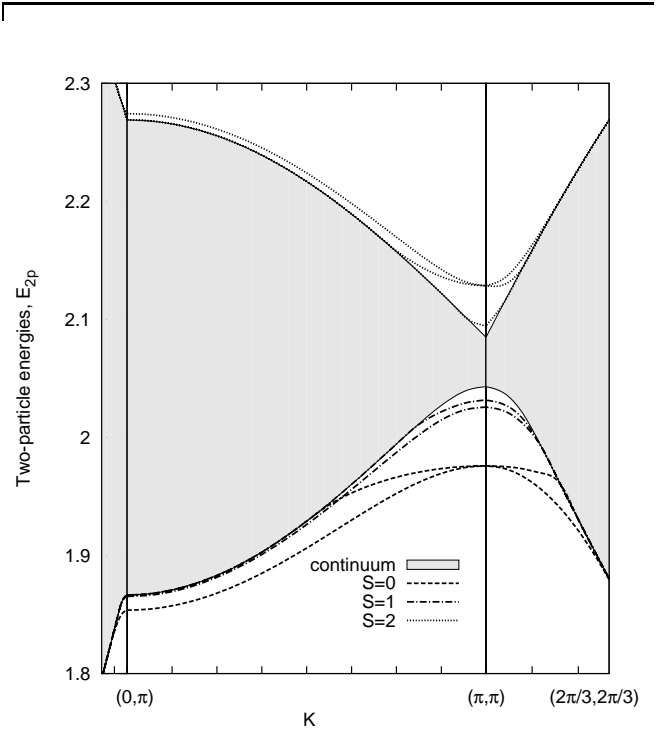}
 \caption{[Color online] Dispersion relations for the two-particle bound/antibound states at $\lambda=0.1$, along symmetry lines
in the Brillouin zone, as calculated from the triplet-wave expansion. 
The 2-particle continuum region is shaded.}
 \label{fig8}
\end{figure}

At leading order in $\lambda$, we find
\begin{eqnarray}
M^S({\bf K},{\bf q},{\bf p}) & \sim &  
-2\lambda[\gamma_{{\bf p+q}}+\gamma_{{\bf p-q}}+\gamma_{{\bf K}/2+{\bf
p}} \nonumber \\
 & & +\gamma_{{\bf K}/2+{\bf q}}+\gamma_{{\bf K}/2-{\bf p}}+\gamma_{{\bf
K}/2-{\bf q}}]
\label{eq56a}
\end{eqnarray}
\begin{equation}
M^T({\bf K,q,p}) \sim 
\lambda[\gamma_{{\bf q+p}}-\gamma_{{\bf q-p}}]
\label{eq56b}
\end{equation}
and
\begin{eqnarray}
M^Q({\bf K,q,p}) & \sim &  
\lambda[\gamma_{{\bf q+p}}+\gamma_{{\bf q-p}}
-2(\gamma_{{\bf K}/2+{\bf
p}} \nonumber \\
 & &  +\gamma_{{\bf K}/2+{\bf q}}+\gamma_{{\bf K}/2-{\bf p}}+\gamma_{{\bf
K}/2-{\bf q}})]
\label{eq56c}
\end{eqnarray}

Then restricting ourselves to the particular momentum ${\bf K} =
(\pi,\pi)$, simple solutions to the Bethe-Salpeter equation (\ref{eq51})
can be found:
\begin{eqnarray}
\Psi^{S,Q}({\bf K,q}) & \sim & (\cos q_x \pm \cos q_y) \nonumber \\
\Psi^{T}({\bf K,q}) & \sim & (\sin q_x \pm \sin q_y) 
\label{eq67}
\end{eqnarray}
corresponding to nearest-neighbour pairs of triplon excitations, with
energies:
\begin{eqnarray}
E^S({\bf K}) & \sim & 2-\lambda \nonumber \\
E^T({\bf K}) & \sim & 2-\lambda/2 \nonumber \\
E^Q({\bf K}) & \sim & 2+\lambda/2
\label{eq68}
\end{eqnarray}
Since the 2-particle continuum is confined strictly to $E_{cont} = 2$ at
this order and this momentum, we see that the singlet and triplet states
are bound states lying below the continuum, while the quintuplet states
are antibound states lying above the continuum. There are two degenerate
states in each case, corresponding to the $\pm$ signs in equation
(\ref{eq67}), or to the two possible axes $x$ and $y$ of the
nearest-neighbour pairs. At higher orders these states may mix and
separate.

These results are easily understood in a qualitative fashion. For an
$S^z = 2$ excitation, for example, the spins on the nearest-neighbour
sites are necessarily aligned parallel, giving rise to a repulsive
interaction; whereas for $S = 0$ or 1 the neighbouring spins can be
aligned either parallel or antiparallel, allowing the possibility of an
attractive interaction.

Solving the wave equation (\ref{eq51}) with vertex functions given by the leading order approximations (\ref{eq52}) 
- (\ref{eq57}), we obtain numerical solutions for the 2-particle spectrum, as illustrated in Figure \ref{fig8}, at a coupling
$\lambda = 0.1$, near momentum ${\bf k} = (\pi,\pi)$. It can be seen that the pairs of degenerate $S=0$ and $S=2$ bound/antibound states
split as one moves away from $(\pi,\pi)$, and all states eventually merge into the continuum. 

\begin{figure}
 \includegraphics[width=1.0\linewidth]{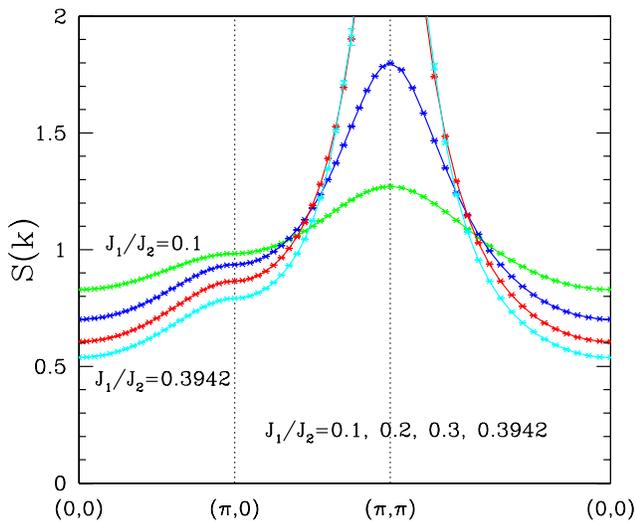}
 \caption{[Color online] 
The total static structure factor $S({\bf k})$ as a function of ${\bf
k}$, for various couplings $\lambda = J_1/J_2$.}
 \label{fig9}
\end{figure}

\section{Series Expansions}
\label{sec3}

We have performed a standard dimer series expansion \cite{singh1988,oitmaa2006} for this
model, where the Hamiltonian is written as
\begin{eqnarray}
H & = & H_0 + \lambda V \\
H_0 & = & \sum_i {\bf S_{1i} \cdot S_{2i}} \\
V & = & \sum_{l=1,2} \sum_{<ij>} {\bf S_{li} \cdot S_{lj}}
\label{eq69}
\end{eqnarray}
and perturbation series are generated for the quantities of interest in
powers of $\lambda$, using linked cluster methods. Details of the linked
cluster approach are reviewed in \cite{oitmaa2006}. In very brief
summary, the ground-state energy per dimer can be written as a sum of
the irreducible contributions (cumulants) coming from every connected
cluster of dimers which can be embedded on the lattice, the order of the
contributions rising with the size of the cluster. The 1-particle
energies can be written in terms of irreducible transition amplitudes $\Delta_1
(i,j)$ of the effective Hamiltonian \cite{gelfand1996}, which consist of a 
sum over all
linked clusters connected to $i$ and $j$, the initial and final
positions of the 1-particle excitations. And finally, the 2-particle
energies can be written in terms of irreducible transition amplitudes $\Delta_2
(i,j;k,l)$ of the 2-particle effective Hamiltonian \cite{trebst2000}, consisting of a sum
over all linked clusters connected to $(i,j)$ and $(k,l)$, the initial
and final positions of the 2-particle excitations. The amplitudes
$\Delta_2$ are then employed in the 2-particle Schr{\" o}dinger or
Bethe-Salpeter equation to calculate the energy for as a function of
momentum. We use a finite-lattice approach \cite{oitmaa2006} for this
purpose, where the Schr{\" o}dinger equation is solved on a finite
lattice in position space, of sufficient size to ensure convergence of
the results.

Once a perturbation series in $\lambda$ has been calculated for a given
quantity, it can be extrapolated to finite $\lambda$ using Pad{\' e}
approximants or integrated differential approximants.

 Zheng \cite{zheng1997}
has previously calculated series for the ground-state energy and
1-particle excitations. These results have already been compared with
the triplet-wave predictions in Figures \ref{fig4}, \ref{fig5} and \ref{fig6}.

\subsection{Structure Factors}
\label{sec3a}

\begin{figure}
 \includegraphics[width=1.0\linewidth]{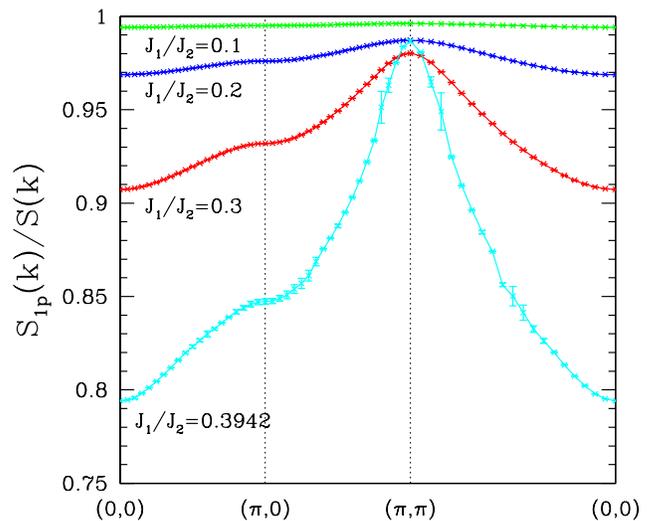}
 \caption{[Color online] 
The ratio $S_{1p}({\bf k})/S({\bf k})$ of the 1-particle static
structure factor to the total static structure factor as a function of
${\bf k}$, for various couplings $\lambda = J_1/J_2$. }
 \label{fig10}
\end{figure}

Figures \ref{fig9} and \ref{fig10} show some series results for
structure factors, which have not been shown before. Figure \ref{fig9}
shows the total static transverse structure factor $S({\bf k}) \equiv S^{+-}({\bf
k})$ as a function of ${\bf k}$ at various couplings $\lambda =
J_1/J_2$, where $S^{+-}({\bf k})$ is the Fourier transform of the
correlation function:
\begin{equation}
S^{+-}({\bf k}) = \frac{1}{N}\sum_{i,j}e^{i{\bf k \cdot
(r_i-r_j)}}<S^+_jS^-_i>_0
\label{eq70}
\end{equation}
\begin{widetext}
\begin{center}
\begin{table}
\caption{Series coefficients of $\lambda^N$ in the expansions for the 1-particle structure factor
$S_{1p}$ and integrated structure factor $S$ at momenta ${\bf k} = (\pi,\pi)$ and $(0,0)$.}
\begin{tabular}{|c|c|c|c|c|}
\hline
 N  &  $S_{1p}(\pi,\pi)$ & $S(\pi,\pi)$ & $S_{1p}(0,0)$ & $S(0,0)$ \\
\tableline
0 & 1.00000000000000D+00 & 1.00000000000000D+00 &  1.00000000000000D+00 &  1.00000000000000D+00 \\ 
1 & 2.00000000000000D+00 & 2.00000000000000D+00 & -2.00000000000000D+00 & -2.00000000000000D+00 \\
2 & 5.00000000000000D+00 & 5.43750000000000D+00 &  3.00000000000000D+00 &  3.43750000000000D+00 \\
3 & 1.20000000000000D+01 & 1.24375000000000D+01 & -7.00000000000000D+00 & -6.56250000000000D+00 \\
4 & 2.60000000000000D+01 & 2.73476562500000D+01 &  1.42500000000000D+01 &  1.48476562500000D+01 \\
5 & 6.19609375000000D+01 & 6.16328125000000D+01 & -3.08359375000000D+01 & -3.09609375000000D+01 \\ 
6 & 1.45859863281250D+02 & 1.46245605468750D+02 &  6.65551757812500D+01 &  6.68159179687500D+01 \\
7 & 3.60063964843752D+02 & 3.57834899902344D+02 & -1.51234863281252D+02 & -1.51278381347656D+02 \\
8 & 8.71365653991730D+02 & 8.80394332885743D+02 &  3.23292167663603D+02 &  3.28300582885742D+02 \\
9 & 2.13146787007666D+03 & 2.15030324554441D+03 & -7.25282606760795D+02 & -7.27275304158507D+02 \\
\hline
\end{tabular}
\label{tab1}
\end{table}
\end{center}
\end{widetext}

All results are for $k_z = \pi$, probing intermediate states antisymmetric between the planes, 
and we only refer to
${\bf k} = (k_x,k_y)$ hereafter.
 
 The dominant feature is a large peak at the N{\' e}el point
${\bf k} = (\pi,\pi)$, which appears to become divergent as $\lambda
\rightarrow \lambda_c$. This behaviour is qualitatively very similar to
that seen in the alternating Heisenberg chain (AHC) in one dimension
\cite{hamer2003}.
 Figure \ref{fig10} shows the ratio of the 1-particle
structure factor $S_{1p}({\bf k})$ to the total $S({\bf k})$ as a
function of ${\bf k}$. The 1-particle contribution generally remains
the dominant part of the total, particularly near the N{\' e}el point.
This behaviour is again reminiscent of the AHC \cite{hamer2003}. 

Further information may be obtained from the series for $S({\bf k})$ and
$S_{1p}({\bf k})$ at the N{\' e}el momentum $(\pi,\pi)$, which are given
in Table I. A Dlog Pad{\' e} analysis of these series, biased at
$\lambda_c =0.3942$, shows both $S({\bf k})$ and
$S_{1p}({\bf k})$ diverging as $\lambda \rightarrow \lambda_c$ with
exponents $-0.68(1)$ and $-0.69(1)$ respectively. The series for the ratio $S_{1p}/S$
shows no sign of a singularity at this point, remaining almost constant, within
2\% of unity at all couplings.
This behaviour is quite different from the AHC case
\cite{affleck1998}, where the ratio vanishes logarithmically at the critical point.

These results should be compared with theoretical expectations. From
scaling theory (see Appendix B),
 the 1-particle structure
factor in the vicinity of the critical point 
$S_{1p}(\pi,\pi)$ should scale like
$(\lambda_c-\lambda)^{(\eta-1)\nu}$,
at the critical (N{\' e}el) momentum.
For the total structure factor at
this point, scaling theory gives exactly the same exponent (see
Appendix B). 
We expect this transition to belong to the universality class of the
O(3) model in 3 dimensions, which has critical exponents
\cite{guida1998}
$\nu = 0.707(4)$, $\eta = 0.036(3)$, hence we expect $(\eta-1)\nu =
-0.682(5)$, which is quite compatible with the numerical estimates
 obtained above. 

How does $S_{1p}$ behave at the critical coupling
away from the N{\' e}el momentum?
In the transverse Ising model \cite{hamer2006}, it was found that the 1-particle residue function
$A({\bf k})$ (see Appendix B) vanishes like $(\lambda_c-\lambda)^{\eta\nu}$ at all momenta,
with a small exponent $\eta\nu = +0.025(3)$, so that
$S_{1p}$ vanishes in the same fashion as $\lambda \rightarrow \lambda_c$. Does the same thing happen
in the present case?  
This is by no means obvious in Figure
\ref{fig10}, which shows the ratio $S_{1p}/S$ dropping slowly as
$\lambda$ increases, but nowhere near zero. 

\begin{widetext}
\begin{center}
\begin{table}
\caption{Series coefficients of $\lambda^N$ for the binding energies in the channels $S = 0,1$, 
and antibinding energy (S = 2). The $S = 0$ and $S = 2$ states are doubly degenerate.
}
\begin{tabular}{|c|c|c|c|c|}
\hline
 N  & S = 0  & S = 1 & S = 1 & S = 2  \\
\tableline
0 &  0.00000000000000D+00 &  0.00000000000000D+00 &  0.00000000000000D+00 &  0.00000000000000D+00 \\
1 &  1.00000000000000D+00 &  5.00000000000000D-01 &  5.00000000000000D-01 &  5.00000000000000D-00  \\
2 & -2.25000000000000D+00 & -2.12500000000000D+00 & -3.12500000000000D+00 & -1.37500000000000D+00  \\
3 & -1.93750000000000D+00 &  1.31250000000000D+00 & -2.93750000000000D+00 &  1.87500000000000D-01  \\
4 & -3.07812500000000D+00 &  2.97656250000002D+00 & -2.77343749999998D+00 &  2.27343750000000D+00  \\
5 &  3.47656250000001D-01 &  1.07812500000003D+00 &  3.06250000000002D+00 &  2.36718750000000D+00  \\
6 & -9.69726562500059D-01 & -1.00527343749999D+01 &  8.35742187500014D+00 & -8.13476562500000D+00  \\
7 &  3.51385498046887D+00 &  7.44207763671879D+00 &  4.07301635742189D+01 & -7.26873779296875D+00 \\
8 & &  7.92327880859462D+00 &  1.69468475341798D+02 & -3.48072814941411D+00   \\
\hline
\end{tabular}
\label{tab2}
\end{table}
\end{center}
\end{widetext}

To pursue this question further, we have studied the series at ${\bf k} = (0,0)$, also given in
Table I. A Dlog Pad{\' e} analysis of these series reveals a dominant singularity at $\lambda =
-0.43(1)$, with exponent around $ -0.65(3)$ in both cases. This will correspond to another
critical point of the model, where the spins order ferromagnetically in the planes, and
antiferromagnetically between them.  At positive $\lambda$, there is
no sign of a pole around $\lambda = 0.4$.
The ratio $S_{1p}/S$ decreases smoothly to around $0.80$ at the critical coupling, 
and shows no sign of vanishing there.
Thus it appears that in this case the renormalized residue function does not vanish
at $\lambda_c$, except at the N{\' e}el momentum. 

\begin{figure}
 \includegraphics[width=1.0\linewidth]{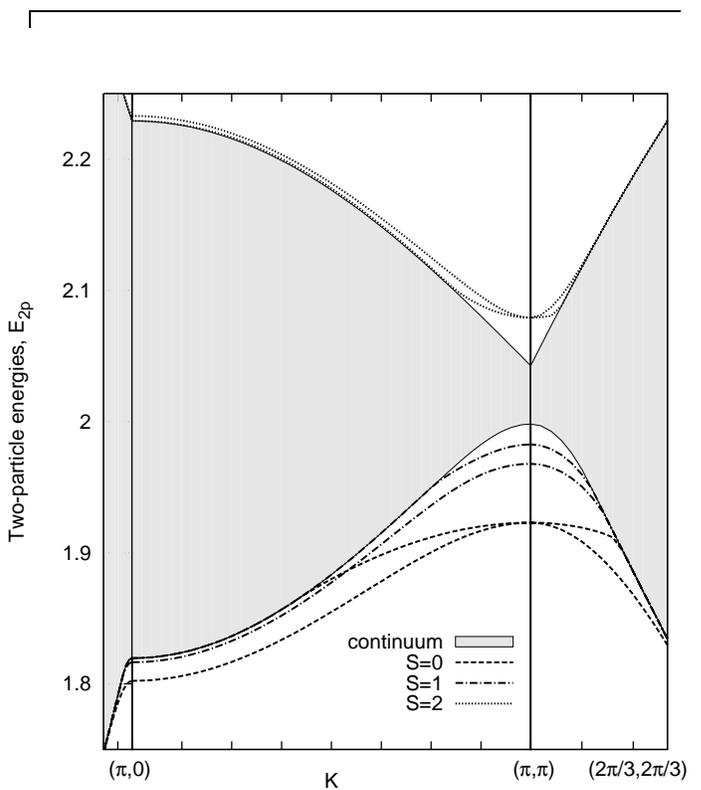}
 \caption{
Series estimates of the energies of 2-particle states at fixed $\lambda = 0.1 $, along 
symmetry lines in the Brillouin zone. 
}
 \label{fig11}
\end{figure}

\subsection{Two-particle excitations}
\label{sec3b}

We have
generalized the computer codes which were previously used to calculate 2-particle
perturbation series for 1-dimensional models \cite{trebst2000} to cover the
two-dimensional case.
Figure \ref{fig11} shows the dispersion diagram estimated from the
perturbation series for 2-particle states at $\lambda = 0.1$.
We have zoomed in on the region where the bound states occur.
 It can be
seen that S = 0 singlet and S = 1 triplet bound states emerge below the
continuum near ${\bf k} = (\pi,\pi)$, and S = 2 quintuplet antibound states appear
above the continuum, as predicted by the triplet-wave theory. 
The $S=0$ and $S=2$ states are
doubly degenerate at ${\bf k} = (\pi,\pi)$. All
states merge with the continuum at some finite momentum point ${\bf k}$, and for
the most part they appear to merge at a tangent, as in the
one-dimensional case \cite{zheng2001}.
The results look very similar to the triplet-wave predictions shown in
Figure \ref{fig8}.

\begin{figure}
\includegraphics[width=1.0\linewidth]{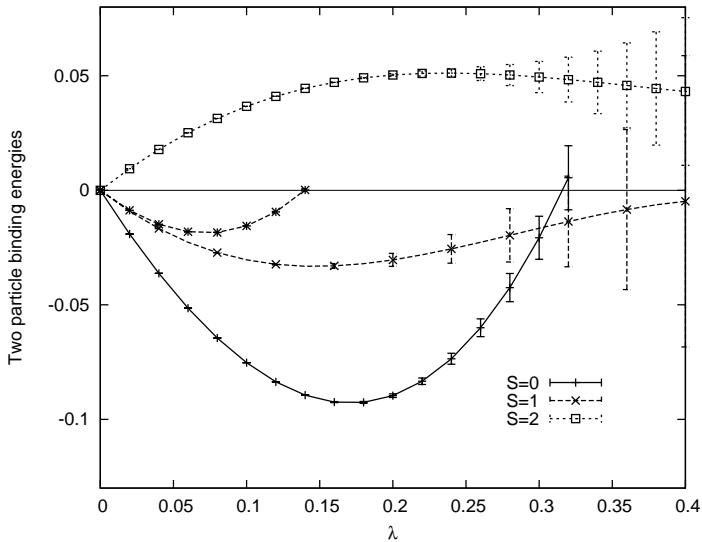}
 \caption{
Binding energies as functions of $\lambda$, relative to the 2-particle continuum.
 For bound states, we graph $(E_{2p}-E^-_{cont})$, while for antibound states we 
graph $(E_{2p}-E^+_{cont})$, where $E^{\pm}_{cont}$ denote the upper/lower bounds of the
continuum.
}
 \label{fig12}
\end{figure}

Figure \ref{fig12} shows the behaviour of the binding energies at ${\bf k}
= (\pi,\pi)$ as functions of $\lambda$, as estimated from Pad{\' e} approximants to the series given in Table \ref{tab2}.
The degenerate pair of singlet bound states are the lowest over most of the range, but merge back into the continuum
somewhat before the critical point. One of the triplet states disappears into the continuum quite early, but the other appears
to disappear only at the critical point.
For the AHC, the binding
energies also vanished at the critical endpoint of the dimerized phase.
The pair of antibound quintuplet states, on the other hand, appear to remain above the continuum even at the
critical point, from our estimates.

\section{Summary and Conclusions}
\label{sec5}

In this paper, we have used a modified triplet-wave theory and dimer series expansions to study the Heisenberg
bilayer system in the dimerized phase. As found in earlier papers \cite{hida1992,zheng1997,sandvik1994}, the model displays a
quantum phase transition from the dimerized phase to a N{\' e}el phase at a coupling ratio $\lambda_c =0.394(1)$,
with critical indices in good agreement with the predicted values from the classical O(3) nonlinear sigma model in
three dimensions,
$\nu = 0.707$ and $\eta = 0.036$.

Our modified triplet-wave approach is found to give good results at small couplings $\lambda$, but towards the
critical region the self-consistent Born approximation approach of Kotov et al. \cite{kotov1998,shevchenko1999}, which
includes some important higher-order terms, gives much better results. The triplet-wave approach predicts, as for
other dimerized systems, two-particle bound states in the $S=0$ and $S=1$ channels where an antiferromagnetic 
alignment of spins can give rise to an attractive force, and antibound states in the $S=2$ channel, where the spin
alignment is necessarily ferromagnetic and repulsive.

Our series calculations focused upon two major features, the critical behaviour of the static transverse structure
factor, and the spectrum of 2-particle bound states in the model. The integrated structure factor $S({\bf k})$ and
the single-particle component $S_{1p}({\bf k})$ were both found to diverge at the critical point for momentum ${\bf k}
= (\pi,\pi)$, with exponents agreeing well with the predicted value $(\eta-1)\nu = -0.68$. The ratio $S_{1p}/S$
remains finite throughout the region, even at the critical coupling $\lambda_c$. This is in contrast to the case of
the alternating Heisenberg chain, where the 1-particle component vanishes logarithmically at the critical point
\cite{affleck1998,hamer2003}. In fact, here the one-particle state dominates everywhere ($S_{1p}/S \ge 80\%$).

In the 2-particle sector, a pair of bound states is found in the $S=0$ and $S=1$ channels near momentum ${\bf k} =
(\pi,\pi)$, as predicted, and a pair
of antibound states in the $S=2$ channel, the pairing being a two-dimensional effect. The singlet $S=0$ states have the
lowest energies at small couplings, but both $S =0$ states and one $S =1 $ state merge back into the continuum as
$\lambda$ increases, leaving only one remaining triplet bound state, which appears to merge with the continuum right
at $\lambda = \lambda_c$. In the S=2 channel, both antibound states appear to remain above the 2-particle continnum
at all couplings $\lambda > 0$.

 As one moves away from ${\bf k} = (\pi,\pi)$, the bound/antibound states eventually
merge into the continuum also. 
They appear to merge with the continuum at a tangent, much as in the
one-dimensional case \cite{hamer2003}.

In future work,
we hope to perform similar calculations for other two-dimensional
models, such as the simple Heisenberg model on the square lattice, and
the Shastry-Sutherland model, which has already been studied by Knetter et
al. \cite{knetter2000}, and where the two-particle states display some intriguing behaviour.

\acknowledgments
This work forms part of a research project supported by a grant
from the Australian Research Council.
We are grateful to the Australian Partnership for Advanced Computing
(APAC) and  to the Australian Centre for Advanced Computing and Communications (ac3) for computational support.

\begin{widetext}
\appendix{}
\begin{center}
{\bf APPENDIX A}
\end{center}

The vertex functions $\Phi^{(i)}_4$ are:

\begin{eqnarray}
\Phi^{(1)}_4(1234) & = & 
\frac{1}{4}[(\gamma_1+\gamma_2)
(c_1c_2+c_1s_2+s_1c_2+s_1s_2)(c_3s_4+s_3c_4)+(\gamma_3+\gamma_4)
(s_1c_2+c_1s_2)(c_3c_4+c_3s_4+s_3c_4+s_3s_4)
 \nonumber \\
& & +\gamma_{{\bf 1+3}}(c_1s_3-s_1c_3)(c_2s_4-s_2c_4)+\gamma_{{\bf
1+4}}(c_1s_4-s_1c_4)(c_2s_3-s_2c_3)]
\label{eqA4}
\end{eqnarray}
\begin{eqnarray}
\Phi^{(2)}_4(1234) & = &
\frac{1}{2}[(\gamma_1+\gamma_2)(s_3c_4+c_3s_4)(c_1c_2+c_1s_2+s_1c_2+s_1s_2)
+(\gamma_3+\gamma_4)(c_1s_2+s_1c_2)(c_3c_4+c_3s_4+s_3c_4+s_3s_4)
\nonumber \\
 & & +\gamma_{{\bf
1-4}}(c_1c_4-s_1s_4)(c_2c_3-s_2s_3)+\gamma_{{\bf
1-3}}(c_1c_3-s_1s_3)(c_2c_4-s_2s_4)]
\label{eqA6}
\end{eqnarray}
\begin{eqnarray}
\Phi^{(3)}_4(1234) & = &
(\gamma_1+\gamma_3)(c_1c_3+c_1s_3+s_1c_3+s_1s_3)(c_2c_4+s_2s_4)
+(\gamma_2+\gamma_4)(c_2c_4+c_2s_4+s_2c_4+s_2s_4)(c_1c_3+s_1s_3)
\nonumber \\
 & & +\gamma_{{\bf
1-4}}(c_1c_4-s_1s_4)(s_2s_3-c_2c_3)+\gamma_{{\bf
1+2}}(c_1s_2-s_1c_2)(s_3c_4-c_3s_4)
\label{eqA7}
\end{eqnarray}
\begin{eqnarray}
\Phi^{(4)}_4(1234) & = &
(\gamma_1+\gamma_2)(c_1c_2+c_1s_2+s_1c_2+s_1s_2)(c_3c_4+s_3s_4)
+(\gamma_3+\gamma_4)(c_1s_2+s_1c_2)(c_3c_4+c_3s_4+s_3c_4+s_3s_4)
\nonumber \\
 & & +\gamma_{{\bf
1-4}}(c_1c_4-s_1s_4)(c_2s_3-s_2c_3)+\gamma_{{\bf
1+3}}(c_2c_4-s_2s_4)(c_1s_3-s_1c_3)
\label{eqA8}
\end{eqnarray}

We have `symmetrized' these expressions with respect to their indices,
using momentum conservation.

\end{widetext}

\appendix{}
\begin{center}
{\bf APPENDIX B. Scaling Theory for Structure Factors}
\end{center}

Let us briefly review scaling theory in the vicinity of a quantum critical point for quantum spin models on a
lattice. Firstly, the integrated or static structure factor \cite{marshall1971,oitmaa2006}
\begin{equation}
S^{\alpha\beta}({\bf k}) = \frac{1}{N} \sum_{i,j} e^{i{\bf k \cdot (r_i - r_j)}} <S^{\alpha}_j S^{\beta}_i>_0
\label{eqB1}
\end{equation}
is just the Fourier transform of the spin correlation function in the ground state, where
$S^{\alpha}_j$ represents the $\alpha$ component of the spin operator at site $j$. In the continuum
approximation near the critical point, this reduces to
\begin{equation}
S^{\alpha\beta}({\bf k}) = \int d^n \ r e^{i{\bf k \cdot r}} <S^{\alpha}({\bf r}) S^{\beta}(0)>_0
\label{eqB2}
\end{equation}
where $n$ is the number of spatial dimensions.

The oscillating factor $\exp(i{\bf k \cdot r})$ will kill off the contributions from large distances
unless it is compensated by a corresponding oscillation $\exp(-i{\bf k_0 \cdot r})$ in the correlation
function. Then we can write
\begin{equation}
S^{\alpha\beta}({\bf k}) = \int d^n r \ e^{i{\bf q \cdot r}} g(r) 
\label{eqB3}
\end{equation}
where ${\bf q} = {\bf k-k_0}$, and g(r) is a smooth function. Scaling theory \cite{cardy1996} then tells us that in
the vicinity of the critical point
\begin{equation}
g(r) \sim r^{-(d-2+\eta)}f(r/\xi)
\label{eqB4}
\end{equation}
where $d=n+1$ is the number of space-time dimensions, and $\xi$ is the correlation length. Thus when
${\bf k} = {\bf k_0}$, the `critical momentum', we have
\begin{eqnarray}
S^{\alpha\beta}({\bf k_0}) & = & \int d^{d-1} r \ r^{-(d-2+\eta)}f(r/\xi) \nonumber \\
 & \sim & \xi^{1-\eta} \int_0^{\infty} d^{d-1}z \ z^{-(d-2+\eta)}f(z) 
\label{eqB5}
\end{eqnarray}
where $z=r/\xi$. As the coupling $\lambda \rightarrow \lambda_c$, corresponding to a quantum phase
transition, we expect
\begin{equation}
\xi \sim (\lambda_c - \lambda)^{-\nu}
\label{eqB6}
\end{equation}
and hence
\begin{equation}
S^{\alpha\beta}({\bf k_0}) \sim (\lambda_c - \lambda)^{-(1-\eta)\nu} , 
\label{eqB7}
\end{equation}
as noted in the text.

For ${\bf q}$ small but non-zero, $|{\bf q}| \ll 1/\xi$, we have
\begin{eqnarray}
S^{\alpha\beta}({\bf k}) 
 & \sim & \xi^{1-\eta} \int_0^{\infty} d^{d-1}z \ z^{-(d-2+\eta)}e^{iqz\xi \cos(\theta)}f(z) 
\nonumber \\
 & \sim & q^{-(1-\eta)} \int_0^{\infty} d^{d-1}z'\ z'^{-(d-2+\eta)}e^{iz' \cos(\theta)}f'(z')
\nonumber \\
 & &
\label{eqB8}
\end{eqnarray}
so that at the critical coupling we expect $S^{\alpha\beta}({\bf k})$ to scale like $q^{-(1-\eta)}$ at
small $q$.

For the 1-particle structure factor, we may paraphrase Sachdev's argument \cite{sachdev1999} as
follows. Assuming relativistic invariance of the effective field theory, which
applies to many though not all models, the dynamic susceptibility in the
vicinity of a quasiparticle
pole is expected to have the form
\bea
\chi ({\bf k},\omega) & = & \frac{A}{c^2{\bf k}^2+\Delta^2-(\omega + i\epsilon)^
2} +
\cdots
\label{eqB12}
\eea
where $\epsilon$ is a positive infinitesimal, $c$ the quasiparticle
velocity,
$\Delta$ is the quasiparticle
energy gap, and $A$ is the ``quasiparticle residue".
Then the dynamic structure factor is
\bea
S ({\bf k},\omega) & = & 
\frac{1}{\pi} Im\{\chi({\bf k},\omega\} 
\label{eqB12a}
\eea

Let
\bea
E({\bf k}) & = & \sqrt{c^2 {\bf k}^2 + \Delta^2}
\label{eqB13}
\eea
then from (\ref{eqB12}), (\ref{eqB12a}) and (\ref{eqB13}) we can write
the dynamic structure factor for the 1-particle state
\bea
S_{\rm 1p} ({\bf k},\omega) & = & 
 \frac{A({\bf k})}{2E({\bf k})} \delta (\omega -
E({\bf k}))
\label{eqB14}
\eea
and hence the static structure factor
\bea
S_{\rm 1p}({\bf k}) & = & 
\int_{-\infty}^{\infty} d\omega S_{1p}({\bf k},\omega)
 = \frac{ A({\bf k})}{2E({\bf k})}
\label{eqB14}
\eea
where $A({\bf k})$ is the residue function.

From renormalization group theory \cite{cardy1996}, 
the scaling dimensions of these quantities are expected to be \cite{hamer2006} 
${\rm dim}[\chi]  =   -2 + \eta$
and 
${\rm dim} [A]  =  \eta$, 
or in other words we expect near the critical point
\bea
A({\bf k}_0) & \sim & (\lambda_c-\lambda)^{\eta \nu}, \nonumber \\
E({\bf k_0}) & \sim & (\lambda_c-\lambda)^{\nu},
\label{eqB15}
\eea
and hence
\bea
S_{1p}({\bf k_0}) & \sim & (\lambda_c-\lambda)^{-(1-\eta)\nu},
\label{eqB16}
\eea
just as for the total structure factor. This is the result quoted in the text.

\end{document}